\documentclass[conference]{IEEEtran}
\IEEEoverridecommandlockouts
\usepackage{cite}
\usepackage{amsmath,amssymb,amsfonts}
\usepackage{algorithmic}
\usepackage{graphicx}
\usepackage{textcomp}
\usepackage{xcolor}

\usepackage{listings,jvlisting}
\usepackage{enumerate}
\usepackage{enumitem}
\usepackage{minted}
\usepackage{caption}

\setminted{linenos, xleftmargin=10pt, numbersep=3pt, fontsize=\scriptsize}

\newcounter{listcounter}
\renewcommand{\thelistcounter}{\arabic{listcounter}}

\newcommand{\jecaption}[2]{%
    \vspace{-0.5em} 
    \refstepcounter{listcounter} 
    \begin{center}
    \footnotesize List \thelistcounter: #1
    \label{#2}
    \end{center}
}

\def\BibTeX{{\rm B\kern-.05em{\sc i\kern-.025em b}\kern-.08em
    T\kern-.1667em\lower.7ex\hbox{E}\kern-.125emX}}
\begin{document}

\title{A Method to Automatically Extract a Network Device Configuration Model by Parsing\\ Network Device Configurations\\}

\author{
    \IEEEauthorblockN{Kosei Nakamura}
    \IEEEauthorblockA{
        \textit{Shinshu University}\\
        Nagano, Japan \\
        24w6052g@shinshu-u.ac.jp
    }
    \and
    \IEEEauthorblockN{Hikofumi Suzuki}
    \IEEEauthorblockA{
        \textit{National Institute of Informatics}\\
        Tokyo, Japan \\
        h-suzuki@nii.ac.jp
    }
    \and
    \IEEEauthorblockN{Shinpei Ogata}
    \IEEEauthorblockA{
        \textit{Shinshu University}\\
        Nagano, Japan \\
        ogata@cs.shinshu-u.ac.jp
    }
    \and
    \IEEEauthorblockN{Hiroaki Hashiura}
    \IEEEauthorblockA{
        \textit{Nippon Institute of Technology}\\
        Saitama, Japan \\
        hashiura@nit.ac.jp
    }
    \and
    \IEEEauthorblockN{Takashi Nagai}
    \IEEEauthorblockA{
        \textit{Institute of Technologists}\\
        Saitama, Japan \\
        t\_nagai@iot.ac.jp
    }
    \and
    \IEEEauthorblockN{Kozo Okano}
    \IEEEauthorblockA{
        \textit{Shinshu University}\\
        Nagano, Japan \\
        okano@cs.shinshu-u.ac.jp
    }
}


\maketitle

\begin{abstract}
When network engineers design a network, they need to verify the validity of their design in a test environment. Since testing on actual equipment is expensive and burdensome for engineers, we have proposed automatic verification methods using simulators and consistency verification methods for a network configuration model. Combining these methods with conventional verification methods for network device configurations will increase the number of verification options that do not require actual devices. However, the burden of writing existing networks into models has been a problem in our model-based verification. In this paper, we propose a method for automatically extracting a network device configuration model by parsing the contents obtained from network devices via show running-config commands and the like. In order to evaluate the effectiveness of the proposed method in realizing round-trip engineering between network device configurations and the network device configuration model, we extracted a model from existing network device configurations and generated device configuration commands. As a result, we obtained model and commands with high accuracy, indicating that the proposed method is effective.
\end{abstract}


\section{Introduction}
When constructing a network, network engineers (hereafter simply referred to as engineers) design the network to meet given requirements\cite{miyata}. The validity of the designed network is then confirmed through testing in a test environment that emulates the production environment\cite{tajima}. However, testing with actual devices poses challenges due to the high financial costs and workload involved in procuring and setting up the equipment. To address these issues, we have previously proposed methods that automatically verify network configurations using simulators without requiring real devices\cite{satake}, as well as methods for automatic consistency verification through static analysis\cite{fujita}, both of which target a network configuration model\cite{arai} that represents the design specifications of the network. Based on the results of these verification processes, we have also proposed a method to reflect corrections made to the network configuration model in the actual device configurations\cite{arai}.
Meanwhile, existing research has proposed lightweight verification approaches that do not rely on physical devices, targeting network device configurations and examining their consistency and correctness\cite{batfish,tajima}. The aspects that can be verified differ across these approaches\cite{satake,fujita,batfish,tajima}. For example, the method proposed by Fujita et al.\cite{fujita} mainly enables the identification of detailed causes of lexical and syntactic errors in device configurations, thereby directly verifying the validity of the configuration descriptions. In contrast, the method proposed by Fogel et al.\cite{batfish} mainly enables simulation analyses of the effects of routing policies and access control lists, thereby verifying the runtime effects of configurations.

In this paper, in order to advance round-trip engineering between network device configurations and network device configuration models via mutual conversion, we propose a method that automatically extracts a network device configuration model by parsing network device configurations. If model-based verification techniques\cite{satake,fujita} and configuration-based verification techniques\cite{batfish,tajima} can be easily combined, it will be possible to conduct verification in a flexible and reciprocal manner: on the one hand, verifying design-phase models without relying on actual or virtual devices, and on the other, performing simulation-based verification of device configurations for operational phases. This will contribute to more comprehensive verification.
To achieve this, existing research has already proposed a method for generating device configuration commands from network configuration models\cite{arai}. However, since methods for converting network device configurations into a network configuration model remain unestablished, engineers must manually convert both device settings and models.This dual approach to technology may impose a significant burden. Therefore, by promoting round-trip engineering through the proposed method, we aim not only to reduce the manual workload of model creation but also to enhance the comprehensiveness of verification.

Here, network device configuration refers to the information obtained from a network device using commands such as show running-config. Furthermore, the network device configuration model refers to a model derived from the network configuration model by excluding connection information, consisting solely of device configuration elements.

To evaluate the effectiveness of the proposed method, we extracted a network device configuration model from existing network device configurations and subsequently generated device configuration commands from the extracted model. As a result, we obtained a correct network device configuration model as well as valid device configuration commands, suggesting that the proposed method is effective in realizing round-trip conversion between network device configurations and the model.

The remainder of this paper is organized as follows. Section 2 explains the technologies addressed in this study. Section 3 describes the proposed method, and Section 4 presents a case study and its results. Section 5 discusses related work, and finally, Section 6 concludes this paper and outlines future work.

\section{Preparation: Introduction to ANTLR}

\begin{figure}[t]
\centering
\includegraphics[width=\columnwidth]{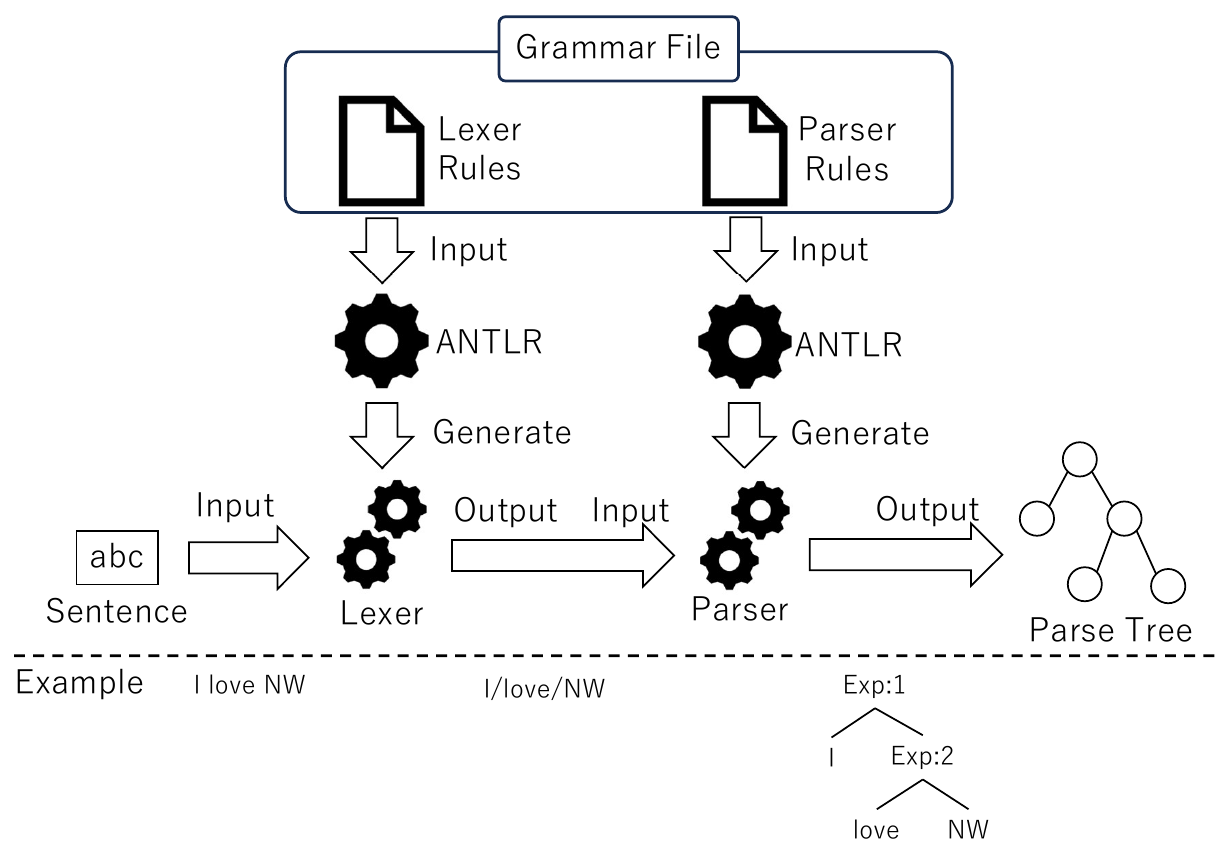}
\caption{Parsing flow using ANTLR}
\label{fig:parse}
\end{figure}

ANTLR\cite{antlr} is a parser generator that can generate parsers for reading, processing, and transforming structured text. Fig.~\ref{fig:parse} shows the Parsing flow using ANTLR. By inputting a grammar file into ANTLR, it generates a lexer and a parser. A grammar file is a file that describes lexer rules and parser rules.
Lexer rules define the tokens (the smallest meaningful units into which text is broken down) that appear in the text.
In ANTLR, these are used to generate the lexer. The lexer
is a program that splits text into tokens. parser rules define the rules determining the structure of sentences and are used by ANTLR to generate a parser. A parser is a program that performs parsing. parsing is the procedure that generates a parse tree according to the parser rules from the sequence of tokens obtained by the lexer. A parse tree is a tree structure representing the structure of the sentence obtained through parsing.
In this research, ANTLR is used as a tool to generate parsers for parsing network device configurations.

\section{Proposed Method} 
\begin{figure}[t]
\centering
\includegraphics[width=\columnwidth]{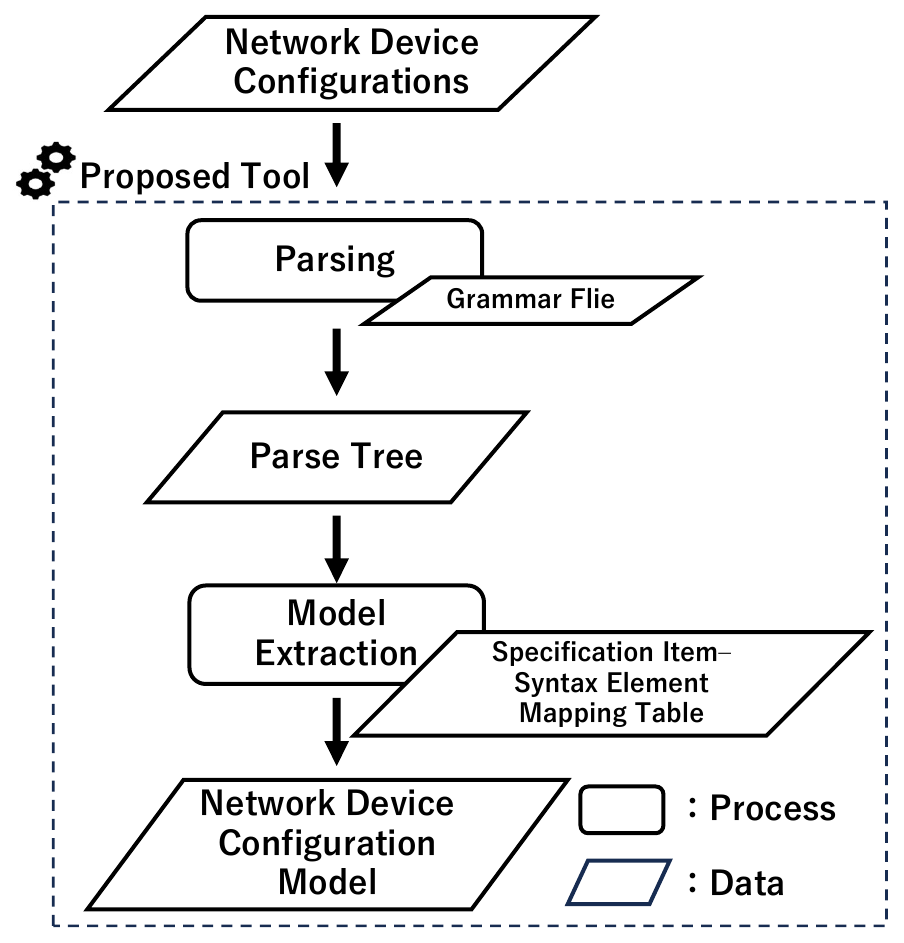}
\caption{Overview of the proposed method}
\label{fig:overview}
\end{figure}

This chapter describes a method for analyzing network device configurations and extracting a network device configuration model. Fig.~\ref{fig:overview} shows an overview of the proposed method. The users of the proposed method are engineers, who require network device configurations, grammar files, and specification item-syntax element mapping tables (hereafter referred to as mapping tables) as input data. In this paper, the network device configuration is provided by the engineer, while the grammar file and correspondence table are provided by the proposed method. This is because the network device configuration depends on the target network being constructed and can only be provided by an engineer, whereas the grammar file and mapping table can be defined once per vendor or device (model) and are reusable. First, we describe the network device configuration input to the proposed method (Section \ref{sec:config}). Next, we describe the network device configuration model, which is the final output of the proposed method (Section \ref{sec:model}). Then, we describe the grammar file for network device configuration provided by the proposed method (Section \ref{sec:grammar}).

The proposed method's procedure overview is as follows.
\begin{enumerate}
\item The proposed method parses the input network device configuration to generate a parse tree (Section \ref{sec:parse}).
\item The proposed method searches the parse tree while referencing the specification item-syntax element mapping table to extract the network device configuration model (Section \ref{sec:extract}).
\end{enumerate}

Finally, the scope and limitations of the proposed method are discussed (Section \ref{sec:limit}).

\begin{figure}[t]
\centering
\begin{lstlisting}[frame=lines, numbers=left]
hostname Router
!
interface FastEthernet3
 switchport access vlan 10
 shutdown
!
interface Vlan10
 ip address 192.168.100.1 255.255.255.0
\end{lstlisting}
\jecaption{show running-config (partial)}{config}
\label{config}
\end{figure}

\subsection{Network Device Configuration} \label{sec:config}
This section describes how to obtain network device configurations, which serve as input to the proposed method. Network device configurations are files describing configuration information obtained via commands from devices such as routers and switches. This research handles network device configurations obtainable via Cisco's show running-config, show version, and show vlan-switch commands, as well as YAMAHA's show config command. List \ref{config} shows a portion of a network device configuration obtained via show running-config. Here, it configures the hostname, assigns IP addresses to VLAN interfaces, and assigns VLANs to switch ports as access ports. Note that the switch port is currently shut down.

\begin{figure}[t]
\centering
\includegraphics[width=\columnwidth]{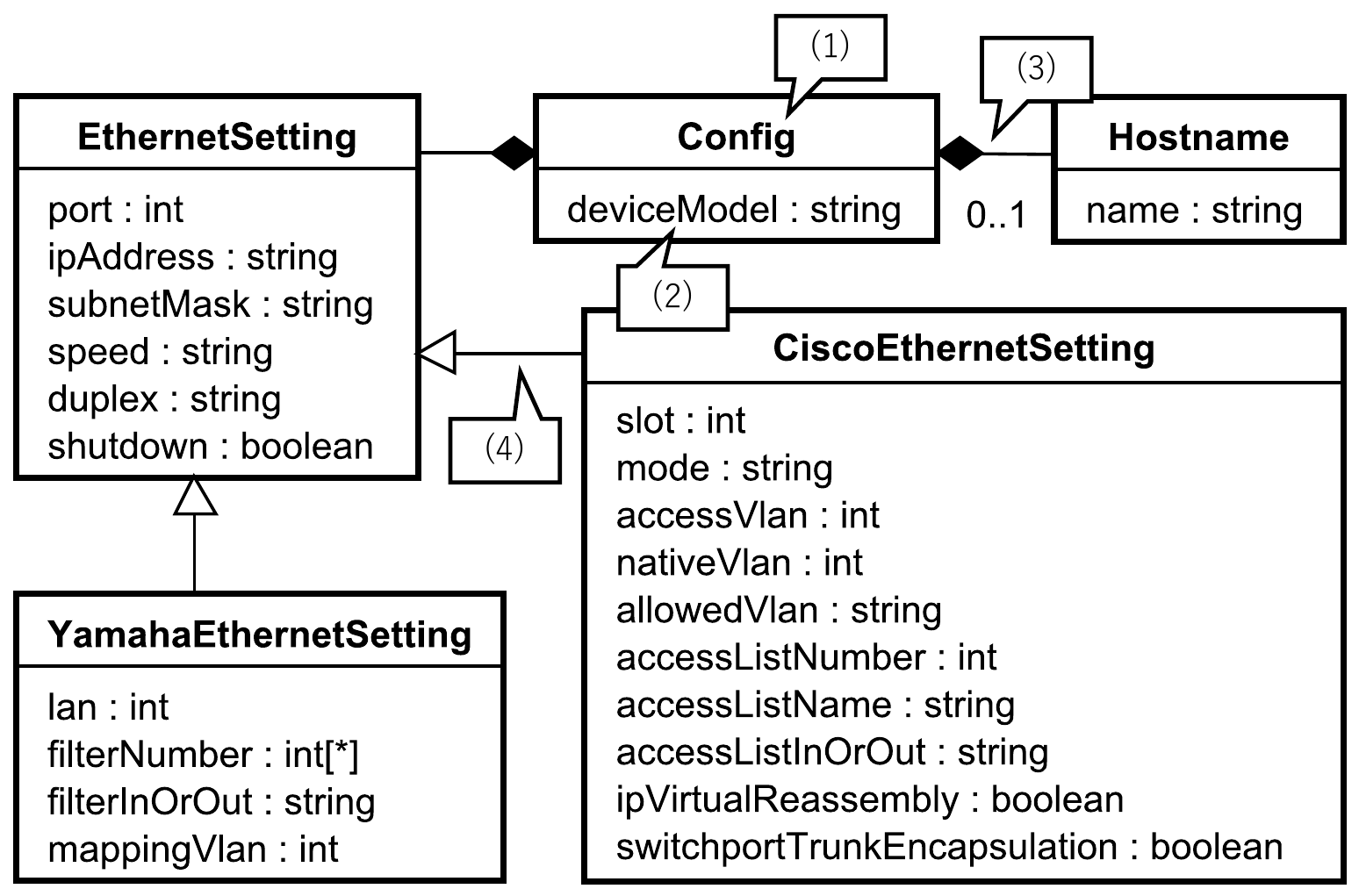}
\caption{Network configuration metamodel}
\label{fig:meta}
\end{figure}

\subsection{Network Device Configuration Model and Its Extensions}\label{sec:model}
The Network Device Configuration Model\cite{arai} is a model based on object diagrams from the object-oriented modeling language UML (Unified Modeling Language), capable of expressing network specifications. Fig.~\ref{fig:meta} shows a network configuration metamodel based on UML class diagram notation, which defines the syntactic structure of the Network Device Configuration Model. The class structure of the network configuration metamodel is organized based on the mode hierarchy of commands used to configure network devices, exhibiting high affinity with network device configurations. The main concepts defining the network configuration metamodel are explained corresponding to the numbers (1) to (4) in the figure. Note that (4) represents an extension from conventional concepts.

\begin{enumerate}
    \item Classify groups of closely related specification items into specification item groups (e.g., Config).
    \item Represent individual items defined within a specification item group as attributes (e.g., deviceModel).
    \item Express relationships between specification item groups using multiplicity and associations (lines connecting specification item groups).
    \item Common specification items across vendors (e.g., ipAddress) are defined in higher-level specification item groups (e.g., EthernetSetting), while vendor-specific items (e.g., accessListNumber) are defined in lower-level groups (e.g., CiscoEthernetSetting), represented through generalization. As an extension of the network configuration metamodel, this research consolidates common specification items for the two vendors involved—Cisco Systems G.K.(hereafter simply Cisco) and Yamaha Corporation (hereafter simply YAMAHA)—into higher-level specification item groups.Specification items appearing only in the device configurations of Cisco or YAMAHA are defined in lower-level specification item groups.
\end{enumerate}

\begin{figure}[t]
\centering
\includegraphics[width=0.7\columnwidth]{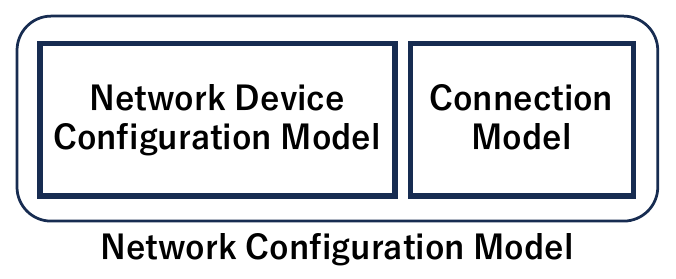}
\caption{Elements of the network configuration model}
\label{fig:composition}
\end{figure}

Fig.~\ref{fig:composition} shows the elements of the network configuration model. The network configuration model is classified into two types: the network device configuration model and the connection model. The network device configuration model represents the configuration of individual devices. The connection model represents the connection information between the ports of each device. This research focuses on extracting the network device configuration model. Note that this classification of network configuration models is an extension of conventional methods.

\begin{figure}[t]
\centering
\includegraphics[width=\columnwidth]{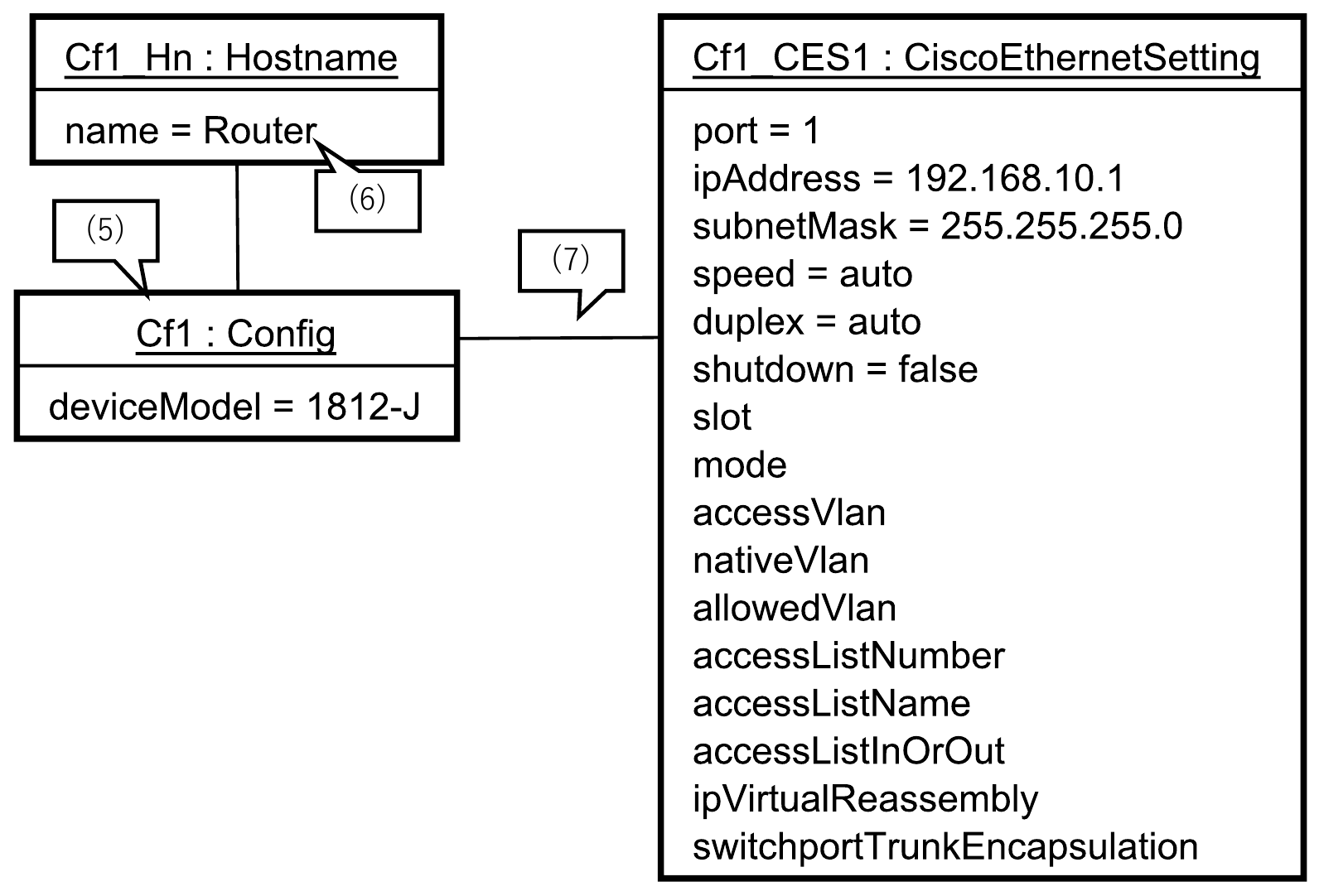}
\caption{Network configuration model}
\label{fig:model}
\end{figure}

Fig.~\ref{fig:model} shows the details of the network device configuration model. Furthermore, the main concepts defining the network device configuration model are explained corresponding to numbers (5) to (7) in the figure.

\begin{enumerate}[resume]
    \item Specification item group values (e.g., Cf1:Config) are represented by instance specifications. The left side of the colon represents the specification item group value name (e.g., Cf1), and the right side represents the specification item group name (e.g., Config). Note that specification item group name values must not duplicate among specification item group values.
    \item Specification item values (e.g., 1812-J) are represented by slot values.
    \item Relationships between specification item group values are represented by links (lines connecting specification item group values).
\end{enumerate}

\subsection{Grammar Files} \label{sec:grammar}
This section describes the grammar files input to the proposed method. This research implemented grammar files\footnote{https://github.com/nakamura-network/grammar} for network device configurations using Cisco's show running-config, show version, and show vlan-switch commands, and YAMAHA's show config command. Part of the lexer rules for List \ref{config} are shown in List \ref{jiku}. In the lexer rules, tokens are described to the left of the colon. To the right of the colon are described strings, regular expressions, tokens, or combinations thereof that fall under this token. An example string is ``interface'', meaning this example is the string ``interface''. An example of a regular expression is [0–9]+, meaning one or more repetitions of any digit from 0 to 9. ? denotes zero or one occurrence. A semicolon indicates the end of a rule definition. Additionally, $|$ (vertical bar) is used to classify different strings, etc., into the same token. For example, lines 9 and 10 indicate that the token MODE\_SETTING is assigned when the string is either ``access'', ``trunk'', ``dynamic auto'', or ``dynamic desirable''. List \ref{kobun} shows the parser rules for List \ref{config}. In parser rules, the sequence of symbols derived from non-terminal symbols (a sequence of one or more non-terminal symbols or terminal symbols) is listed as production rules. A non-terminal symbol is a symbol appearing on the left-hand side (left of the colon) of any production rule, representing a symbol that can be replaced by another sequence of symbols. A terminal symbol is a symbol appearing only on the right-hand side (right of the colon) of a production rule, which is not further transformed by the production rule; in this paper, tokens correspond to this category. In parser rules, non-terminal symbols are written on the left-hand side, and patterns composed of non-terminal symbols or tokens are written on the right-hand side. When defining non-terminal symbols or tokens consecutively on the right-hand side, a space character is inserted between symbols. For example, in the production rule on line 16, the non-terminal symbol access\_vlan is derived into the terminal symbol MODE\_SETTING, followed by the non-terminal symbol vlan\_num. According to line 14, the non-terminal symbol vlan\_num is derived into the token VLAN, followed by the token NUM.

\begin{figure}[t]
\centering
\begin{lstlisting}[frame=lines, numbers=left]
INTERFACE:'interface';
ETHERNET:'FastEthernet'|'GigabitEthernet'
|'TenGigabitEthernet';
HOSTNAME:'hostname';
IP:'ip';
ADDRESS:'address';
SHUTDOWN:'shutdown';
SWITCHPORT:'switchport';
NO:'no';
MODE_SETTING:'access'|'trunk'|'dynamic auto'
|'dynamic desirable';
IF_VLAN:'Vlan';
VLAN:'vlan';
IP_ADDRESS_NUM:NUM'.'NUM'.'NUM'.'NUM;
NUM:[0-9]+;
CHAR:[a-zA-Z]+;
\end{lstlisting}
\jecaption{Lexical rules (partial)}{jiku}
\label{jiku}
\end{figure}

\begin{figure}[t]
\centering
\begin{lstlisting}[frame=lines, numbers=left]
file:category+;
category:interface|(omitted);
hostname:HOSTNAME any;
interface:INTERFACE interface_name;
interface_name:bri|ethernet|if_vlan;
ethernet:
ETHERNET ((stack'/')?slot'/')?port interface_setting;
interface_setting:
(ip_address|switchport|SHUTDOWN|(omitted))*;
if_vlan:IF_VLAN NUM interface_setting;
if_ip_address:NO? IP ADDRESS ip_address subnet_mask;
ip_address:IP_ADDRESS_NUM;
subnet_mask:IP_ADDRESS_NUM;
switchport_setting:
SWITCHPORT(port_mode|access_vlan|trunk);
access_vlan:MODE_SETTING vlan_num;
vlan_num:VLAN NUM;
port:NUM;
any:(NUM|CHAR)+;
\end{lstlisting}
\jecaption{Syntactic rules (partial)}{kobun}
\end{figure}

\subsection{Parsing} \label{sec:parse}
\begin{figure*}[t]
\centering
\includegraphics[width=0.9\linewidth]{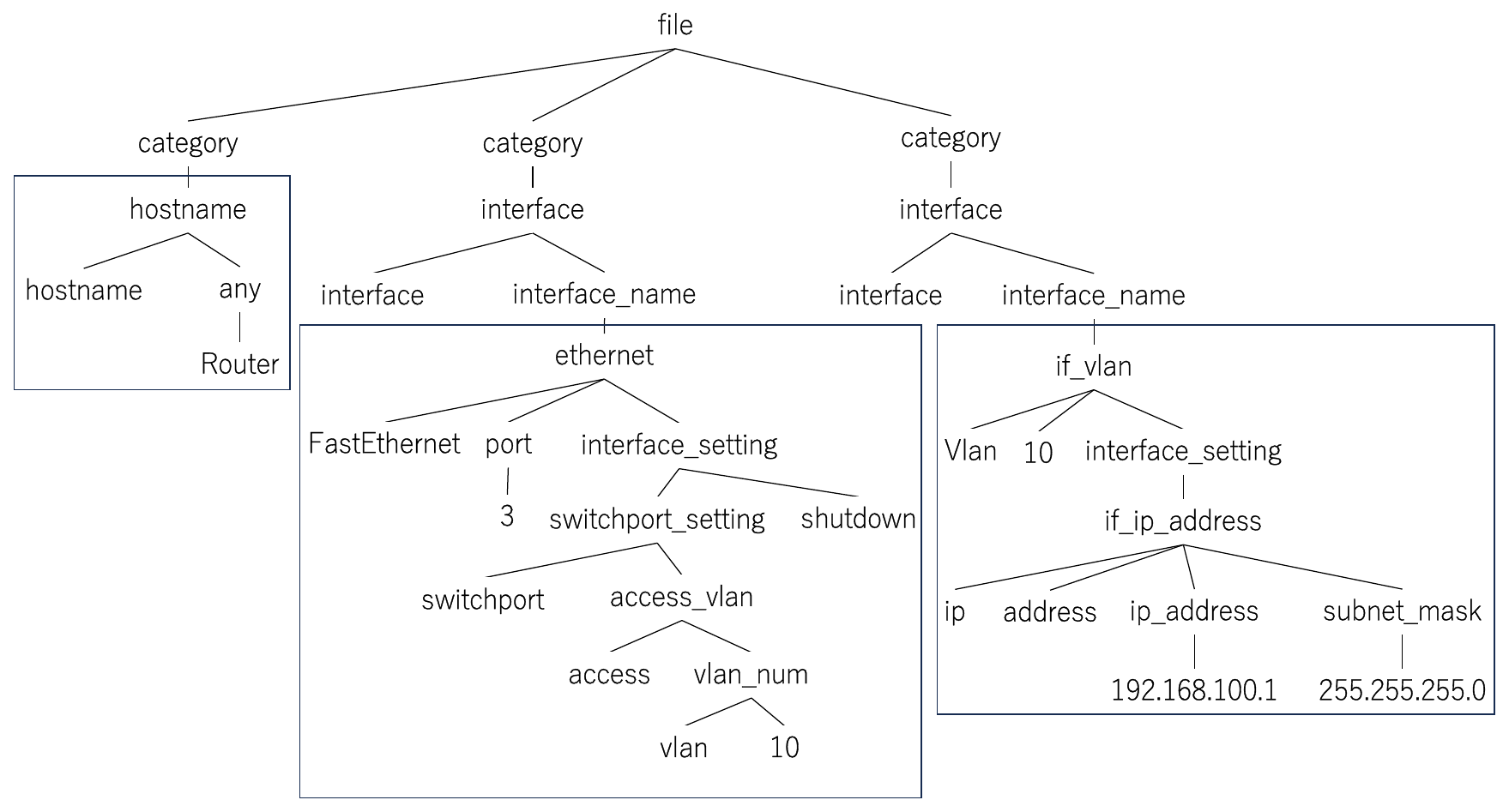}
\caption{Parse tree generated from the network device configuration in list\ref{config}}
\label{fig:AST}
\end{figure*}

This section describes the parsing of network device configurations and the resulting parse tree. Parsing is performed and parse trees are generated using a parser created from the lexer rules and parser rules shown in List \ref{jiku} and List \ref{kobun}, respectively. Fig.~\ref{fig:AST} shows the parse tree generated by parsing the network device configuration shown in List \ref{config}. As an example, for the text ``ip address 192.168.100.1 255.255.255.0'', ``ip'' is tokenized as IP (line 5 of List \ref{jiku}), ``address'' as ADDRESS (line 6 of List \ref{jiku}), and splitting ``192.168.100.1'' and ``255.255.255.0'' into tokens IP\_ADDRESS\_NUM (row 14 in List \ref{jiku}). The syntax tree is then generated according to the production rule if\_ip\_address (row 11 in List \ref{kobun}).

\begin{table*}[t]
  \centering
  \caption{Specification Item–Syntax Element Mapping Table in Cisco (partial)}
  \scalebox{0.9}{
  \begin{tabular}{lll|lll|ll}
    \hline\hline
    \multicolumn{3}{c|}{\textbf{Parse Tree}} & \multicolumn{3}{c}{\textbf{Processing Rule}} & \multicolumn{2}{|c}{\textbf{Network Configuration Metamodel}}\\
    \textbf{Root of Subtree} & \textbf{Parent of Target Node} & \textbf{Target Node} & \textbf{Node Presence} & \textbf{Original} & \textbf{Replaced} & \textbf{Specification Item Group} & \textbf{Specification Item}\\
    \hline
    hostname & hostname & any & Present & & & Hostname & name\\
    ethernet & port & NUM & Present & & & CiscoEthernetSetting & port\\
    ethernet & interface\_setting & SHUTDOWN & Present & .+ & true & CiscoEthernetSetting & shutdown\\
    ethernet & interface\_setting & SHUTDOWN & Absent & .* & false & CiscoEthernetSetting & shutdown\\
    ethernet & access\_vlan & MODE\_SETTING & Present & & & CiscoEthernetSetting & mode\\
    ethernet & vlan\_num & NUM & Present & & & CiscoEthernetSetting & accessVlan\\
    if\_vlan & if\_vlan & NUM & Present & & & CiscoVlanSetting & vlanNum\\
    if\_vlan & interface\_setting & SHUTDOWN & Present & .+ & true & CiscoVlanSetting & shutdown\\
    if\_vlan & interface\_setting & SHUTDOWN & Absent & .* & false & CiscoVlanSetting & shutdown\\
    if\_vlan & ip\_address & IP\_ADDRESS\_NUM & Present & & & CiscoVlanSetting & ipAddress\\
    if\_vlan & subnet\_mask & IP\_ADDRESS\_NUM & Present & & & CiscoVlanSetting & subnetMask\\
    \hline
  \end{tabular}
  }
  \label{tab:rule}
\end{table*}

\begin{figure*}[htb]
\centering
\includegraphics[width=\linewidth]{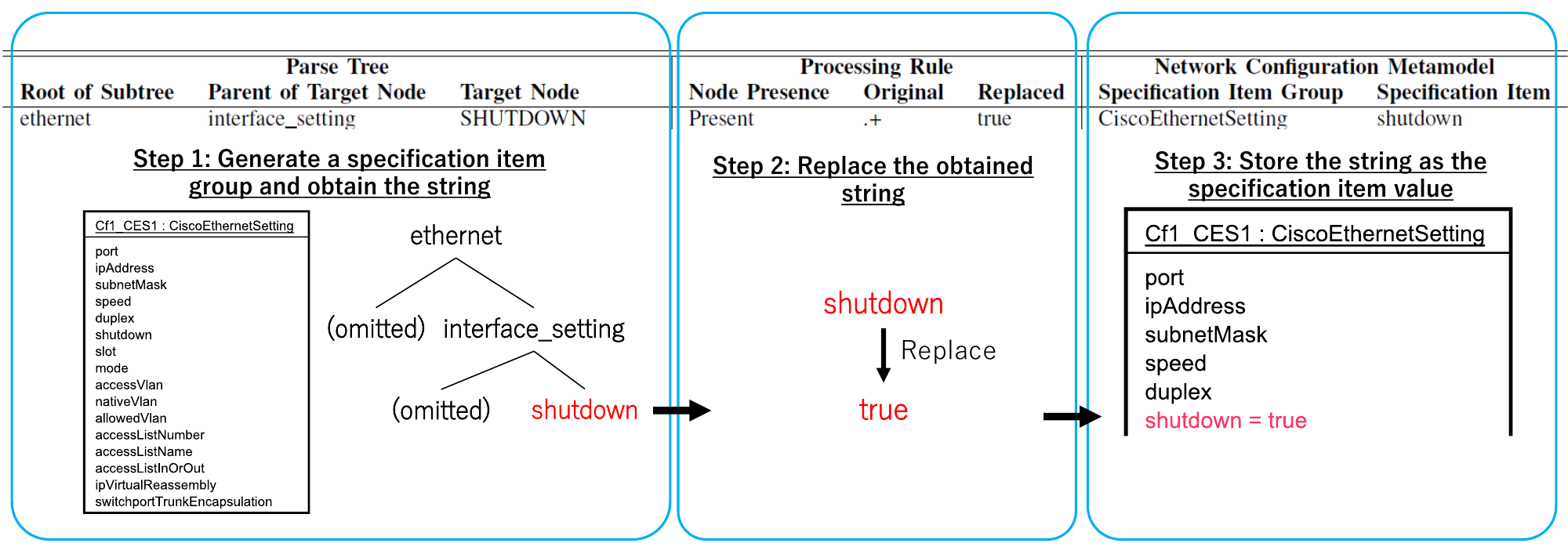}
\caption{Example of model extraction}
\label{fig:extract}
\end{figure*}

\subsection{Model Extraction}\label{sec:extract}
For the parse tree generated by parsing, we perform string extraction and replacement processing related to device configurations using a depth-first search in the forward direction. The proposed method establishes a mapping table that maps elements of the parse tree to elements of the network device configuration model and specifies the replacement processing content. TABLE~\ref{tab:rule} shows the mapping table for Cisco. The “Parse Tree” column in the mapping table represents information for identifying the parse tree element to be replaced. The “Processing Rule” column represents the replacement processing applied to the identified parse tree element. The “Network Configuration MetaModel” column represents information identifying the model element to which the substitution is made. 

First, the substitution procedure is described below and is repeated until all elements are searched from the root node of the parse tree.

\begin{enumerate}
    \item When reaching the symbol $r$ in the ``Root of Subtree'' column of the mapping table, generate the specification item group value $g$ for the ``Specification Item Group'' corresponding to $r$. Subsequently, continue searching beyond $r$. When reaching the symbol $p$ in the ``Parent of Target Node'' column where $r$ is the ``Subtree Root'' column value, verify whether the symbol $t$ in the ``Target Node'' column exists directly under $p$. For example, in step 1 of Fig.~\ref{fig:extract}, when reaching ``ethernet'', the specification item group value for ``CiscoEthernetSetting'' is generated. Then, when reaching ``interface\_setting'' during the search after ``ethernet'', check whether the token SHUTDOWN exists. If the token SHUTDOWN is not found by the end of the search in the subtree under ``interface\_setting'', it is determined that the token SHUTDOWN does not exist. Note that ``shutdown'' in Fig.~\ref{fig:extract} represents the string denoted by the SHUTDOWN token.
    \item Replace $t$ according to the ``Processing Rule'' column in TABLE~\ref{tab:rule}. The ``Node Presence'' column serves as a conditional branch based on $t$'s existence. ``Present'' indicates processing when t exists, while ``Absent'' indicates processing when $t$ does not exist. The ``Original'' column represents the part $l$ of the string represented by $t$ that is to be replaced using a regular expression. If this column is empty, the string represented by $t$ is not replaced. The ``Replaced'' column represents the string that replaces $l$. For example, in step 2 of Fig.~\ref{fig:extract}, since the token SHUTDOWN exists, the processing in the third row of TABLE~\ref{tab:rule} is selected. The regular expression ``.+'' in the ``Original'' column means ``one or more characters.'' Since the replacement string is ``true,'' the string ``shutdown'' represented by the token SHUTDOWN is replaced with ``true.''
    \item Refer to the ``Network Configuration Metamodel'' column in TABLE~\ref{tab:rule} and store the replaced string in the specification item value. The storage location is the specification item corresponding to the target node's parent $p$ within the specification items where the ``Specification Item Group'' is $g$. For example, in step 3 of Fig.~\ref{fig:extract}, the ``true'' obtained in the previous step is stored in the specification item value ``shutdown'' of the specification item group value ``CiscoEthernetSetting''.
\end{enumerate}

\begin{figure}[t]
\centering
\includegraphics[width=\columnwidth]{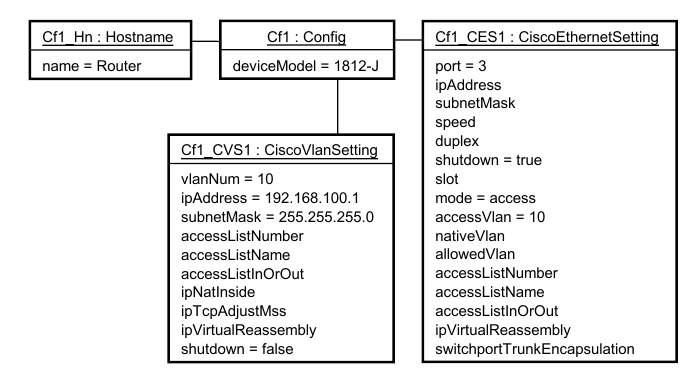}
\caption{Network device configuration model generated from list\ref{config}}
\label{fig:configmodel}
\end{figure}

\begin{figure*}[t]
\centering
\includegraphics[width=0.8\linewidth]{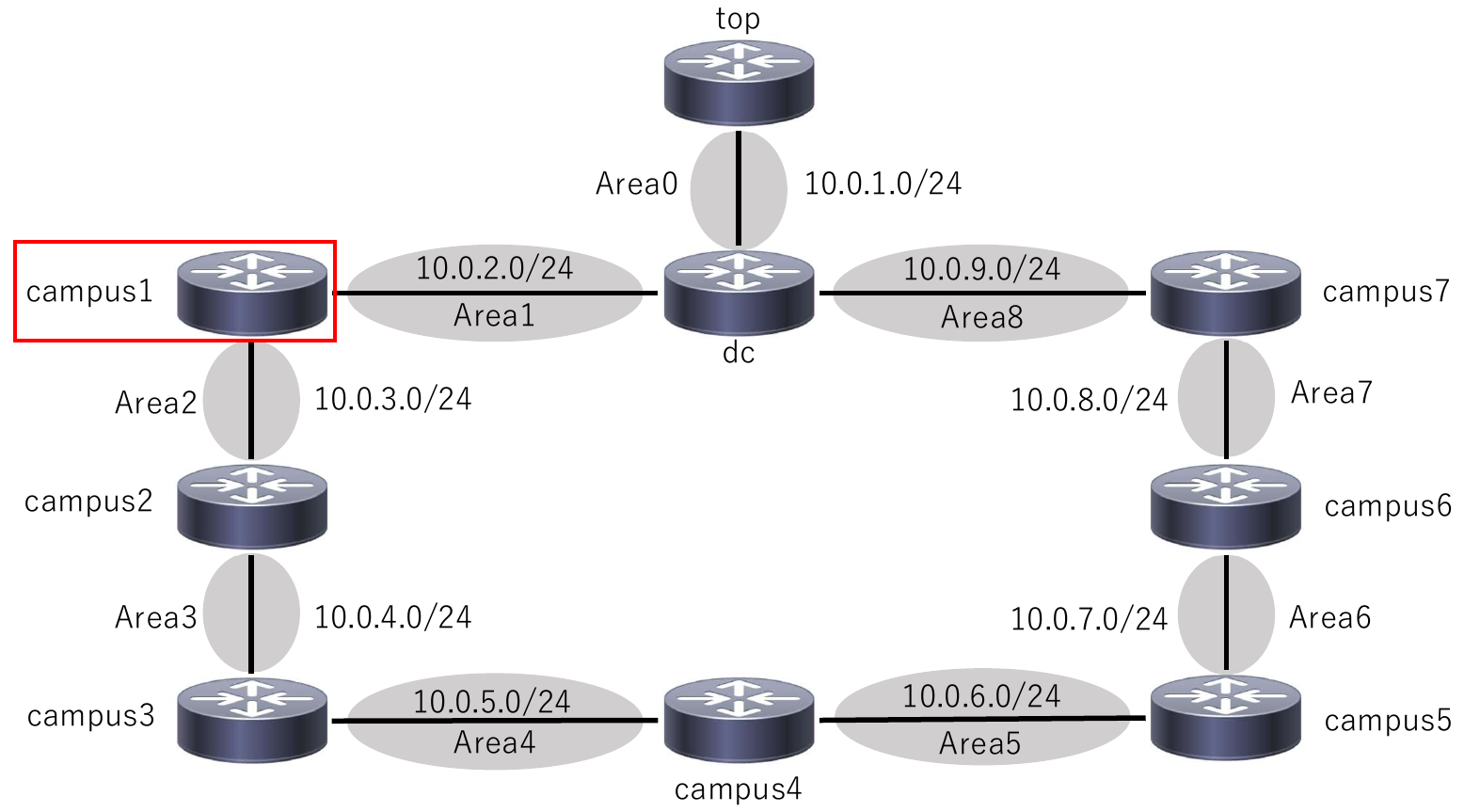}
\caption{Network of application example}
\label{fig:ospf}
\end{figure*}

Next, we describe how to generate links between specification item group values. When reaching the symbol r1 contained in the ``Subtree Root'' column of TABLE~\ref{tab:rule}, the specification item group value generated here is called $g_1$. Furthermore, during the search of the subtree rooted at $r_1$, when reaching the symbol $r_2$ contained in the ``Subtree Root'' column, the specification item group value generated here is called $g_2$. Then, a link is generated between $g_1$ and $g_2$. In this manner, links are generated based on the inclusion relationships of subtrees detected from the root node of the syntax tree until the exploration is complete. Note that during the exploration of the root ``file'' node of the syntax tree, the specification item group value for the specification item group Config is generated. For example, in the subtree search of Fig.~\ref{fig:AST}, the specification item group value ``Config'' is first generated during the search of the ``file'' node. Upon reaching the ``hostname'' node, the specification item group value ``Hostname'' is generated. A link is then created between ``Config'' and ``Hostname''. Fig.~\ref{fig:configmodel} shows the network device configuration model extracted from the device settings in List \ref{config}. The black-bordered area in the syntax tree of Fig.~\ref{fig:AST} represents the target subtrees for extraction. By exploring the subtree rooted at the ``hostname'' node, the subtree rooted at the ``ethernet'' node, and the subtree rooted at the ``if\_vlan'' node, the specification group values ``Hostname'', ``CiscoEthernetSetting'', and ``CiscoVlanSetting'' are created. ``CiscoVlanSetting'' are created, and links are generated between these and the specification item group value ``Config''. Furthermore, the substituted values obtained from the syntax tree are set for each specification item value.

\begin{figure*}[t]
\centering
\includegraphics[width=\linewidth]{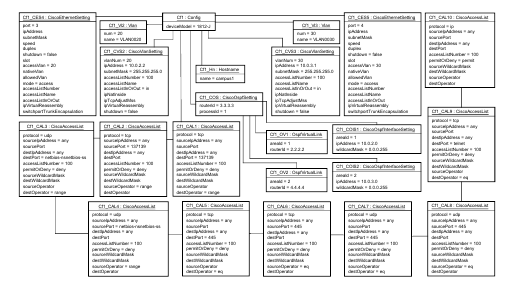}
\caption{network device configuration model (campus1)}
\label{fig:ospf_model}
\end{figure*}

\subsection{Scope of This Proposal}\label{sec:limit}
In round-trip engineering, ideally, the network device configuration model should comprehensively cover all configuration items of the target network devices, and all their configuration values should be extracted correctly. Conversely, it should also be possible to comprehensively and correctly convert from the network device configuration model to the target device's configuration. However, the scope of this paper's proposal is limited, focusing only on the device configurations covered by the preceding methods\cite{arai,satake,fujita}. The specific scope of the proposal is shown below, and extraction from these has been achieved.

\begin{enumerate}
    \item Hostname setting
    \item Ethernet settings
    \item VLAN settings
    \item VLAN interface settings
    \item Static route settings
    \item STP(Spanning Tree Protocol) settings
    \item OSPF (Open Shortest Path First) settings
    \item ACL (Access Control List) settings (Filter settings)
\end{enumerate}

Previous approaches\cite{satake,fujita} listed interface settings (including Ethernet and VLAN) and STP settings as verification items. Therefore, within the current proposal scope, we have successfully extracted the device configurations used for verification in previous approaches\cite{satake,fujita}. The proposed method currently targets the aforementioned device configurations, defining its scope while considering consistency with existing specifications. Therefore, future expansion will involve exploring the generalization of the model to support a broader range of vendors and operating systems.

\section{Case Study}
To evaluate the effectiveness of the proposed method, we investigated whether it could correctly extract a network device configuration model from an existing network. We verified the accuracy of the extracting network device configuration model using the mapping table for the application case\cite{nakamura}. Since this research builds upon prior work\cite{nakamura}, the same application case as in\cite{nakamura} was used. To investigate whether mutual conversion between the model and device configurations is possible, the extracted network device configuration model was applied to the prior method\cite{arai}, and the accuracy of the generated device configuration commands was evaluated.

\subsection{Application Case}
Fig.~\ref{fig:ospf} shows the topology of the existing network used as the application case\cite{nakamura}. This network mimics the Shinshu University network and employs the OSPF routing protocol. The devices used are Cisco892J (top, dc) and Cisco1812J (campus1 to campus7). 

\subsection{Procedure}
The case study procedure is outlined below.
\begin{enumerate}
    \item Input network device configurations from the application example into the proposed method to extract the network device configuration model.
    \item Verify the correctness of the extracted network device configuration model.
    \item Input the extracted network device configuration model into the prior method\cite{arai} to output network device configuration commands.
    \item Applied the output network device configuration commands to the actual devices and verified the correctness of the configuration.
\end{enumerate}

\subsection{Results}
For steps (1) and (2), we describe the results of applying the proposed method to the case study network. From the extracted network device configuration model, we extracted all 1102 specification item values set during network construction. We confirmed that strings were correctly replaced according to the processing rules defined in the mapping table. Furthermore, we measured the number of distinct types of given specification items (e.g., the left-hand side of vlanNum = 20) for the extracted specification item values. This yielded 32 distinct types. Since the total number of specification item types defined by the network device configuration model is 66, this confirms the correct extraction of specification item values for approximately 48.5\% of all specification items. Regarding the generation of links between specification item group values, we confirmed that links matching the relationships defined in the network configuration metamodel were drawn. As an example, Fig.~\ref{fig:ospf_model} shows a portion of the network device configuration model corresponding to the device (campus1) within the red box in Fig.~\ref{fig:ospf}. 

\begin{figure}[t]
\centering
\begin{lstlisting}[frame=lines, numbers=left]
enable
configure terminal
router ospf 1
router-id 3.3.3.3
area 1 virtual-link 2.2.2.2
area 2 virtual-link 4.4.4.4
network 10.0.2.0 0.0.0.255 area 1
network 10.0.3.0 0.0.0.255 area 2
exit
interface fastethernet 3
no shutdown
switchport mode access
switchport access vlan 20
exit
vlan 30
name VLAN0030
exit
interface vlan 30
ip address 10.0.3.1 255.255.255.0
ip access-group 100 in
exit
interface fastethernet 4
no shutdown
switchport mode access
switchport access vlan 30
exit
interface vlan 20
ip address 10.0.2.2 255.255.255.0
ip access-group 100 in
exit
hostname campus1
vlan 20
name VLAN0020
exit
exit
copy running-config startup-config
\end{lstlisting}
\jecaption{A Result of applying the previous method\cite{arai}}{command}
\end{figure}

Regarding steps (3) and (4), we describe the results of applying the extracted network device configuration model to the prior method\cite{arai}. As output from the prior method\cite{arai}, device configuration commands were generated, excluding 10 ACL-related settings. Furthermore, when the generated commands were applied to the actual device, it was confirmed that the commands were correctly configured on the device. The results of applying the prior method\cite{arai} are shown in List \ref{command}.

\subsection{Discussion}
\subsubsection{Extraction of the Network Configuration Model}
The case study demonstrated the successful extraction of the network device configuration model, indicating the proposed method is likely effective. In the prior method\cite{satake}, wiring information is required when configuring links for each device on the network simulator. Similarly, prior method\cite{fujita} also requires wiring information for consistency checks between multiple nodes and for certain protocol-dependent verifications. Therefore, we plan to explore automatically generating wiring information specification item groups by analyzing the LLDP (Link Layer Discovery Protocol) protocol as an extension of the proposed method. Furthermore, the number of specification item values in the network device configuration model extracted from the applied example network exceeded 1000, requiring considerable time to verify the model's correctness. Therefore, the prior method\cite{fujita}, which can automatically point out the validity of configuration values for the model, is considered effective for more complex network device configuration models. Furthermore, while we confirmed that specification item values were correctly extracted for 32 types of specification items, verifying the correctness of extraction for the remaining 34 types remains a future task. The unconfirmed specification items mainly include static route settings and STP (Spanning Tree Protocol).

\begin{table*}[t]
  \centering
  \caption{Characteristic differences between network configuration model-based verification and device-configuration-based verification}
  \begin{tabular}{l|l|l}
    \hline\hline
    \textbf{Characteristic differences} & \textbf{Fujita et al.\cite{fujita}} & \textbf{Fogel et al.\cite{batfish}}\\
    \hline
    Analysis of incomplete configurations & $\circ$ & $\triangle$\\
    Strict detection of lexical / syntactic errors & $\circ$ & $\triangle$\\
    CLI-style output of network design information & $\circ$ & ×\\
    Verification of routing protocol behavior & × & $\circ$\\
    Verification of security policy enforcement & × & $\circ$\\
    \hline
  \end{tabular}
  \\\small $\circ$: supported,\; $\triangle$: partially supported,\; ×: not supported
  \label{tab:diff}
\end{table*}

\subsubsection{Application to the Previous Method}
Except for ACL-related settings, we confirmed that device configuration commands were generated correctly. Therefore, we believe network management is possible through the iterative process of extracting network configuration models from device settings, verification using the network configuration model as input, and generating device configuration commands. The failure to generate device configuration commands for ACLs was caused by the previous method\cite{arai} referenced the empty specification item value for the specification item group value “CiscoAccessList,” leading to the determination that the specification item values required for command generation were insufficient. In Cisco ACL configuration, settings are possible even without including elements like port number specifications in the command. Therefore, the prior method\cite{arai} also needs to be extended to match the proposed method, enabling command generation even when some specification item values are empty. Furthermore, the prior method\cite{arai} generates commands using a graph search approach targeting a network configuration metamodel adhering to Cisco's command hierarchy. Consequently, the generated device configuration commands vary depending on the graph search order. This could potentially cause issues with device configuration commands for other vendors. For example, YAMAHA device configuration commands lack the command hierarchy found in Cisco but have configurations with constraints on command input order, potentially causing issues with those configurations. Therefore, sufficient performance verification of the proposed method remains a future challenge.

\section{Related Work}
Regarding network information extraction and validity verification through network device configuration analysis, Fogel et al.\cite{batfish} proposed a tool (Batfish) that performs automated verification via syntactic parsing of network device configurations. Batfish models the control plane and data plane based on the parsed device configurations to verify configuration consistency. 
TABLE~\ref{tab:diff} shows the characteristic differences between the network configuration model-based verification method by Fujita et al.\cite{fujita} and the device configuration-based verification approach by Fogel et al.\cite{batfish}. Differences in verification items and suitable verification scenarios are evident. Utilizing the generation of device configuration commands\cite{arai} to enable concurrent use of the proposed method and methods\cite{fujita,batfish} for device configurations is expected to enhance verification capabilities.

Regarding efforts for centralized network management, NAPALM (Network Automation and Programmability Abstraction Layer with Multivendor support)\cite{napalm} is a Python library for managing network device configurations with multivendor support. NAPALM enables connecting to, retrieving, and manipulating device configurations using vendor-independent, unified commands. NAPALM provides technology for operational phases such as retrieving and modifying device configurations and is not intended for verification. Conversely, the proposed method aims to enhance verification during the design phase. Therefore, we believe there is a clear distinction in the intended use cases for each technology.

Regarding efforts to model network configurations, OpenConfig\cite{open_config} works to standardize network management by using vendor-neutral data models defined in the YANG language and network configuration control protocols. OpenConfig primarily focuses on improving operational management efficiency. It defines common configuration items and includes a mechanism to extend vendor-specific configurations, enabling unified configuration management. However, OpenConfig's scope is primarily operational management. For example, it cannot immediately reflect new vendor-specific configurations and is not necessarily suitable for detailed configuration definition and verification during the design phase. The network device configuration model addressed in this research focuses on defining and verifying network configurations during the design phase. It differs from OpenConfig in its approach, specifically by not directly standardizing differences in configuration items between vendors. Instead, it adopts a flexible extension method applying object-oriented design inheritance, allowing configuration item differences to coexist.

The proposed method improves upon our previous approach\cite{nakamura}. The previous method could not handle cases where specification items with the same name existed in different specification item groups; the proposed method addresses this. Furthermore, while the network configuration model had been extended to include ACLs since the previous study\cite{nakamura}, we confirmed that the proposed method can still correctly extract the network device configuration model.

\section{Conclusion}
This paper proposed an automatic extraction method for network device configuration models through syntax analysis of network device configurations, aiming to realize round-trip between configurations and models. Case study results indicate that the proposed method can reliably extract network configuration models from device settings. Furthermore, applying the extracted models to the prior method\cite{arai} successfully generated correct network configuration commands. This facilitates the adoption of prior method\cite{satake} and enables flexible network design through bidirectional conversion between models and device configurations.

The practice identified in this research involves discovering, through case studies, the approaches and challenges involved in uniformly representing device configurations from different vendors using a network device configuration model. This practice will contribute to streamlining the mutual conversion between models and device configurations when enhancing verification methods in the future by combining model-based verification techniques\cite{fujita} with configuration-based verification techniques\cite{batfish}. As identified approaches, first, when modeling device configurations across different vendors, we proposed a method for unified handling across vendors. This involves abstracting settings with common meaning (e.g., subnet mask) by standardizing them at the specification item group level, while allowing differences in display formats through specification item group values, thereby avoiding unnecessary model bloat. Furthermore, to flexibly accommodate new configuration items added by each vendor, we demonstrated a model extension method using UML inheritance. This allows vendor-specific configuration items to be defined as subclasses, enabling coexistence between different vendors. Even when type definitions are coexistent, the advantage lies in obtaining the minimal necessary model by selecting only the subclasses required for the target network and instantiating them (creating specification item group values). However, challenges remained: flexibility was needed in defining grammar files to accommodate differences in device configurations across vendors, and a mechanism was required to reduce the burden of mapping grammar files to network configuration models. 

By accumulating improvements to address these challenges, the proposed method enables the early detection of misconfigurations during the design phase and establishes a consistent configuration definition and verification process for environments with mixed devices from different vendors. Specifically, by applying the method of Fujita et al.\cite{fujita} to establish a process for early detection of misconfigurations during the design phase and for consistent configuration definition and verification in environments with mixed vendors. In particular, verifying the consistency of network configurations, such as conflicts in ACLs or routing settings, is expected to reduce problems before actual deployment. Furthermore, combining this approach with network configuration-based verification methods like Batfish\cite{batfish} enables integrated verification from both design and operational perspectives, enhancing practicality for network engineers. Advancements in such verification methods are expected to streamline the mutual conversion between this research's model and device configurations, contributing to improved verification workflows in network design and operations.

Future challenges include exploring methods to expand the meta-model to support more multi-vendor environments and implementing automated device configuration verification techniques by applying prior methods\cite{satake}. Furthermore, future prospects include investigating model-driven device migration support that accounts for configuration differences between vendors and developing methods to automate the mapping between syntax elements and specification items using generative AI.

We extend our deepest gratitude to Yamaha Corporation for their cooperation in this research.

\bibliographystyle{IEEEtran}
\bibliography{mybib}

\end{document}